\documentstyle [12pt]{article} \textwidth 155mm \textheight 220mm

\begin{document}
\begin{center}
{\bf Thermal Entanglement between Alternate Qubits of a Four-qubit\\
Heisenberg $XX$ Chain in a Magnetic Field }
\end{center}

\begin{center}
\vskip0.2cm

Min Cao and Shiqun Zhu*

\vskip0.2cm {\it
School of Physical Science and Technology,\\
Suzhou University, Suzhou, Jiangsu 215006,\\
 People's Republic of China}
\end{center}

\bigskip

\begin{abstract}
The concurrence of two alternate qubits in a four-qubit Heisenberg
$XX$ chain is investigated when a uniform magnetic field $B$ is
included. It is found that there is no thermal entanglement
between alternate qubits if $B$ is close to zero. Magnetic field
can induce entanglement in a certain range both for the
antiferromagnetic and ferromagnetic cases. Near zero temperature,
the entanglement undergoes two sudden changes with increasing
value of the magnetic field $B$. This is due to the changes in the
ground state. This novel property may be used as quantum
entanglement switch. The anisotropy in the system can also induce
the entanglement between two alternate qubits.\\

\noindent PACS\#: 03.65.Ud, 03.67.Lx,75.10.Jm.\\

\bigskip

\noindent {*}Corresponding author, E-mail: szhu@suda.edu.cn

\end{abstract}

\newpage

The existence of entanglement shows many interesting properties in
quantum systems. Its nonlocal quantum correlation has become one
of the most valuable resources in quantum communication [1-3] and
quantum computation [4, 5]. Recently, the concept of thermal
entanglement in solids was introduced and studied in
one-dimensional anisotropic Heisenberg model [6]. Thermal
entanglement in two-qubit Heisenberg $XX$ chain was investigated
with and without an external magnetic field [7-9]. The
entanglement between qubits of the next and next-next neighbors in
an open spin chain and in multi-qubit Heisenberg model was
presented [10-13].

In most of the previous investigations, the concurrence of two
nearest-neighbor qubits is calculated as a measure of
entanglement. For a pair of two qubits, the concurrence is given
by [14,15]
\begin{equation}
\label{eq1} C = \max \{ {\lambda _1 - \lambda _2 - \lambda _3 -
\lambda _4 ,0}\}
\end{equation}

\noindent where the quantities $\lambda_i (i=1, 2, 3, 4)$ are the
square roots of the eigenvalues of the operator
\begin{equation}
\label{eq2} \varrho = \rho_{12}(\sigma_1^y \otimes
\sigma_2^y)\rho_{12}^\ast (\sigma_1^y \otimes \sigma_2^y)
\end{equation}
in descending order. The values of the concurrence are ranged from
zero to one when quantum states are changed for unentangled to
maximally entangled states.

The state of the system at thermal equilibrium is represented by
the density operator
\begin{equation}
\label{eq3}\rho (T) = \frac{1}{Z}\exp ( { - \frac{H}{kT}} )
\end{equation}

\noindent where $Z=Tr[exp(-H/kT)]$ is the partition function, $k$
is the Boltzmann's constant and $T$ is the temperature. Since
$\rho(T)$ represents a thermal state, the entanglement in the
state is called thermal entanglement [16-17].

In this report, a Heisenberg $XX$ model of four-qubit in a linear
chain is investigated when a magnetic field $B$ is included. The
pairwise entanglement between alternate qubit is calculated. The
four-qubit $XXM$ Heisenberg model is described by the Hamiltonian
\begin{equation}
\label{eq4} H_{XXM} = J\sum\limits_{n = 1}^4 {( {\sigma _n^ +
\sigma _{n + 1}^ - + \sigma _n^ - \sigma _{n + 1}^ + })} +
B\sum\limits_{n = 1}^4 {\sigma _n^z }
\end{equation}

\noindent where $\sigma _n^\pm $ are the raising and lowering
operations, $B$ is the external magnetic field and perpendicular
to the chain, $J$ is the strength of interaction. The value of
positive and negative $J$ corresponds to the antiferromagnetic and
ferromagnetic cases respectively.

The eigenvalues and eigenstates of the Hamiltonian in Eq. (4) can
be calculated analytically. In a four-qubit Heisenberg chain with
periodic boundary conditions, the eigenvalues are given by
\begin{equation}
\label{eq5}
\begin{array}{l}
E_0 = - 4B, \qquad \qquad \quad  E_1 = 2J - 2B, \qquad \quad E_2 = E_4 = - 2B \\
E_3 = - 2J - 2B, \qquad \quad E_5 = 2\sqrt 2 J, \qquad \qquad E_6 = - 2\sqrt 2 J \\
E_7 = E_8 = E_9 = E_{10} = 0,\qquad \qquad \qquad \qquad E_{11} = 2J + 2B    \\
E_{12} =E_{14} = 2B,\qquad \quad E_{13} = - 2J + 2B, \quad E_{15} = 4B\\
\end{array}
\end{equation}

\noindent and the corresponding eigenstates can be explicitly
expressed as
\begin{equation}
\label{eq6}
\begin{array}{l}
\left| {\psi _0 } \right\rangle = \left| {0000} \right\rangle \\
\left| {\psi _1 } \right\rangle = \frac{\displaystyle
1}{\displaystyle 2}\left( {\left| {0001} \right\rangle + \left|
{0010} \right\rangle + \left| {0100} \right\rangle +
\left| {1000} \right\rangle } \right)\\
\left| {\psi _2 } \right\rangle = \frac{\displaystyle
1}{\displaystyle 2}\left( {\left| {0001} \right\rangle + i\left|
{0010} \right\rangle - \left| {0100} \right\rangle -
i\left| {1000} \right\rangle } \right) \\
\left| {\psi _3 } \right\rangle = \frac{\displaystyle
1}{\displaystyle 2}\left( {\left| {0001} \right\rangle - \left|
{0010} \right\rangle + \left| {0100} \right\rangle -
\left| {1000} \right\rangle } \right) \\
\left| {\psi _4 } \right\rangle = \frac{\displaystyle
1}{\displaystyle 2}\left( {\left| {0001} \right\rangle - i\left|
{0010} \right\rangle - \left| {0100} \right\rangle +
i\left| {1000} \right\rangle } \right)\\
\left| {\psi _5 } \right\rangle = \frac{\displaystyle \sqrt 2
}{\displaystyle 4}\left( {\left| {0011} \right\rangle + \left|
{0110} \right\rangle + \left| {1100} \right\rangle + \left| {1001}
\right\rangle } \right) + \frac{\displaystyle 1}{\displaystyle
2}\left( {\left| {0101}
\right\rangle + \left| {1010} \right\rangle } \right)\\
\left| {\psi _6 } \right\rangle = \frac{\displaystyle \sqrt 2
}{\displaystyle 4}\left( {\left| {0011} \right\rangle + \left|
{0110} \right\rangle + \left| {1100} \right\rangle + \left| {1001}
\right\rangle } \right) - \frac{\displaystyle 1}{\displaystyle
2}\left( {\left| {0101}
\right\rangle + \left| {1010} \right\rangle } \right)\\
\left| {\psi _7 } \right\rangle = \frac{\displaystyle
1}{\displaystyle 2}\left( {\left| {0011} \right\rangle + i\left|
{0110} \right\rangle - \left| {1100} \right\rangle -
i\left| {1001} \right\rangle } \right)\\
\left| {\psi _8 } \right\rangle = \frac{\displaystyle
1}{\displaystyle 2}\left( {\left| {0011} \right\rangle - \left|
{0110} \right\rangle + \left| {1100} \right\rangle -
\left| {1001} \right\rangle } \right)\\
\left| {\psi _9 } \right\rangle = \frac{\displaystyle
1}{\displaystyle \sqrt 2 }\left( {\left| {0101}
\right\rangle - \left| {1010} \right\rangle } \right)\\
\left| {\psi _{10} } \right\rangle = \frac{\displaystyle
1}{\displaystyle 2}\left( {\left| {0011} \right\rangle - i\left|
{0110} \right\rangle - \left| {1100} \right\rangle +
i\left| {1001} \right\rangle } \right)\\
\left| {\psi _{11} } \right\rangle = \frac{\displaystyle
1}{\displaystyle 2}\left( {\left| {1110} \right\rangle + \left|
{1101} \right\rangle + \left| {1011} \right\rangle +
\left| {0111} \right\rangle } \right)\\
\left| {\psi _{12} } \right\rangle = \frac{\displaystyle
1}{\displaystyle 2}\left( {\left| {1110} \right\rangle + i\left|
{1101} \right\rangle - \left| {1011} \right\rangle -
i\left| {0111} \right\rangle } \right)\\
\left| {\psi _{13} } \right\rangle = \frac{\displaystyle
1}{\displaystyle 2}\left( {\left| {1110} \right\rangle - \left|
{1101} \right\rangle + \left| {1011} \right\rangle -
\left| {0111} \right\rangle } \right)\\
\left| {\psi _{14} } \right\rangle = \frac{\displaystyle
1}{\displaystyle 2}\left( {\left| {1110} \right\rangle - i\left|
{1101} \right\rangle - \left| {1011} \right\rangle +
i\left| {0111} \right\rangle } \right)\\
\left| {\psi _{15} } \right\rangle = \left| {1111} \right\rangle\\
\end{array}
\end{equation}

If the concurrence of two alternate qubits is considered, the
reduced density matrix $\rho_{13}=Tr_{24}\rho(T)$ can be given by
\begin{equation}
\label{eq7}
 \rho _{13} \left( T \right) = \frac{\displaystyle 1}{\displaystyle Z}\left( {{\begin{array}{*{20}c}
 u \hfill & 0 \hfill & 0 \hfill & 0 \hfill \\
 0 \hfill & w \hfill & y \hfill & 0 \hfill \\
 0 \hfill & y \hfill & w \hfill & 0 \hfill \\
 0 \hfill & 0 \hfill & 0 \hfill & v \hfill \\
\end{array} }} \right) \\
\end{equation}

\noindent in the basis $\{\left| {00} \right\rangle, \left| {01}
\right\rangle, \left| {10} \right\rangle, \left| {11}
\right\rangle \}$. Where $u, v, y$ are
\begin{equation}
\label{eq8}
\begin{array}{l}
u = \frac{\displaystyle1}{\displaystyle2}\left( 1+{e^{\left( {2J +
2B} \right)\beta } + e^{ - \left( {2J - 2B}
\right)\beta }} + cosh2\sqrt 2 J\beta \right) + e^{2B\beta } + e^{4B\beta } \\
v =  \frac{\displaystyle 1}{\displaystyle 2}\left( 1+{e^{ - \left(
{2J + 2B} \right)\beta } + e^{\left( {2J - 2B} \right)\beta }} +
cosh2\sqrt 2 J\beta  \right)+e^{ - 2B\beta } + e^{ - 4B\beta } \\
y = \frac{\displaystyle 1}{\displaystyle 2}\left(-1+cosh\left( {2J
+ 2B} \right)\beta + cosh\left( {2J - 2B} \right)\beta +
cosh2\sqrt 2 J\beta \right)- cosh2B\beta  \\
\end{array}
\end{equation}

\noindent The partition function of the system is
\begin{equation}
\label{eq9} Z = 4 \left( 1+cosh2B\beta \right) + 2 \left[
cosh4B\beta + cosh\left( {2J + 2B} \right)\beta + cosh\left( {2J -
2B} \right)\beta + cosh2\sqrt 2 J\beta \right]
\end{equation}

\noindent with $\beta = \frac{\displaystyle 1}{\displaystyle {k
T}}$. In the following calculations $k$ is set to be $1.0$.

From the Eqs. (1), (2) and (7), the concurrence can be obtained
\begin{equation}
\label{eq10} C = \frac{2}{Z}\max \left( {\left| y \right| -
\sqrt{uv},0} \right)
\end{equation}

Except the concurrence $C$ of two qubits, the global entanglement
$Q$ of many-qubit pure states also needs to be considered. The
global entanglement $Q$ is introduced by [18-20]

\begin{equation}
\label{eq11} Q=\frac1N\sum_{i=1}^NIC_i^2
\end{equation}
where the $i$-concurrence of $IC_i$ means the entanglement between
the qubit $i$ and the other qubits and can be expressed as

\begin{equation}
\label{eq12} IC_i=\sqrt{2[1-Tr(\rho_i^2)]}
\end{equation}

The concurrence $C$ as functions of the magnetic field $B$ and the
temperature $T$ is plotted in Fig. 1. The strength of interaction
$J$ is chosen to be $1.0$. Fig. 1(a) is a three-dimensional plot
of the concurrence $C$ as functions of $B$ and $T$. From Fig.
1(a), it is clear that there is a two-peak structure in $C$. The
two peaks appear symmetrically at two sides of $B$. There is no
entanglement between alternate qubits at $B=0$. The entanglement
is also independent of $T$ [10]. This can be understood since
$B=0$ the ground state will be $\left| {\psi _6 } \right\rangle$,
which is unentangled between alternate qubits. When the
temperature $T=0$, the ground state energy as a function of the
magnetic field $B$ is changed from $E_6$ to $E_3$, and then to
$E_{0}$ for antiferromagnetic case. While the ground state energy
is changed from $E_5$ to $E_{11}$, and then to $E_{15}$ for
ferromagnetic case. Therefore there should be a change in the
concurrences $C$ because of the change in the ground state. It can
be seen that the increase of magnetic field $B$ cannot induce
entanglement when the temperature $T$ is very high. When the
temperature $T$ is very low, the entanglement is increased with
increasing value of the magnetic field $|B|$ to a maximum value.
Then it is decreased and finally disappeared. The maximum value of
the entanglement decreases when the temperature $T$ is increased.

The contour map of the concurrence $C$ as functions of the
magnetic field $B$ and the temperature $T$ is plotted in Fig.
1(b). Four contours of $C=0.5, 0.3, 0.1, 0$ are shown
respectively. Beyond the contour $C=0$, the entanglement is equal
to zero. It means that there exists a critical temperature $T_c$.
From the curve of $C=0$ in Fig. 1(b), it can be seen that the
critical temperature $T_c$ depends on the magnetic field $B$. If
$B<0.09$, the concurrence $C$ is always zero no matter the
temperature $T$ is increased or decreased. When $0.09<B<0.41$ or
$B>1.0$, there are two critical temperatures of $T_c$. When $T$ is
either above the lower part of the curve of $C=0$ or below the
upper part of $C=0$, the concurrence $C$ is always greater than
zero. When $0.41<B<1.0$, there is a single value of the critical
temperature $T_c$. When the temperature $T$ is above the curve of
$C=0$, the entanglement is vanished no matter the magnetic field
$B$ is increased or decreased. For four-qubit nearest-neighbor
$XXM$ model, the critical temperature is independent of the
magnetic field $B$ [11].

The thermal entanglement of alternate qubits for $J=-1.0$ is also
studied. The result is almost the same as that shown in Fig. 1.
This means that the entanglement exits for both antiferromagnetic
and ferromagnetic cases.

The concurrence $C$, the global entanglement $Q$, and the
$i$-concurrence $IC_i$ are plotted in Fig. 2 as a function of the
magnetic field $B$. The concurrence $C$ of different qubits is
plotted in Figs. 2(a) and 2(b) when the temperature $T$ is varied
with $T=0.01$, $0.1$, and $0.5$. Fig. 2(a) is a plot of $C$ for
alternate qubits. From Fig. 2(a), it can be seen that the shape of
$C$ is like a square wave for low temperature of $T=0.01$. The
concurrence $C$ keeps zero until $B$ is increased to $0.41$. Then
the concurrence maintains a maximal value of $C=0.5$ until it
drops to zero at $B=1.0$. In the limit of $T\rightarrow 0$, one
has

\begin{equation}
\label{eq13}
\begin{array}{l}
\mathop {\lim }\limits_{T \to 0} C =0 \qquad\qquad\quad\qquad\quad
\quad \quad
|B| < (\sqrt 2-1) |J| \\
\mathop {\lim }\limits_{T \to 0} C = \frac{1}{2}\qquad \qquad
\qquad \quad \quad
(\sqrt 2-1) |J|\leq |B| \leq |J| \\
\mathop {\lim }\limits_{T \to 0} C = 0       \qquad\qquad
\qquad\qquad\qquad \quad \quad
|B| > |J| \\
\end{array}
\end{equation}

\noindent This can be understood as follows. When $|B|
> |J|$ and $|B| < (\sqrt 2-1) |J|$, the ground states are the
unentangled state $\left| {\psi _0 } \right\rangle$ and $\left|
{\psi _6 } \right\rangle$ respectively. While for $(\sqrt 2-1)
|J|<|B|<|J|$, the ground state is the maximally entangled state
$\left| {\psi _3 } \right\rangle$ of the two alternate qubits.
When the temperature $T$ is increased to $0.1$, and $0.5$, the
shape of $C$ is changed from square wave to a single peak. The
maximal value of the concurrence $C$ is decreased when $T$ is
increased. For finite temperatures, this definite ground state
structure is smoothed out by the partition of higher states and
therefore the concurrence $C$ becomes smaller with higher
temperatures. To compare this with that of the two
nearest-neighbor qubits in the four-qubit $XXM$ model, the
concurrence $C$ of the two nearest-neighbor qubits is plotted as a
function of $B$ in Fig. 2(b) with the same condition. For low
temperature of $T=0.01$, the entanglement keeps a constant value
of $C=0.46$ until it drops to a dip. It seems that the dip is due
to the crossing of energy level at the point of $B=(\sqrt 2-1)$
[11]. Then $C$ maintains a maximal value of $C=0.5$ until it drops
to zero at $B=1.0$. When the temperature $T$ is increased to
$0.1$, the shape of $C$ is changed to two peaks with almost the
same dip. When $T$ is increased to $0.5$, the dip disappears.
There is only a single peak at $B=0$. From Figs. 2(a) and 2(b), it
can be seen that the effects of temperature on $C$ is much
stronger for alternate qubits than that for nearest qubits. When
$T$ is increased to $0.5$, the curve of $C$ in Fig. 2(b) is much
higher than that in Fig. 2(a). It seems that the temperature
affects the entanglement between weakly interacting alternate
qubits stronger than that for strongly interacting nearest qubits.
The global entanglement $Q$ and the $i$-concurrence $IC_i$ are
plotted in Fig. 2(c). Due to the symmetry of the eigenstates in
Eq. (6), the values of $i$-concurrence are the same. When $0\leq
B\leq0.41$, $IC_1=IC_2=IC_3=IC_4=1.0$. The value of
$i$-concurrence is the same as that of $Q$. The two-qubit
concurrence $C$ is $C_{12}=C_{14}=C_{23}=C_{34}=(2\sqrt2-1)/4$ and
$C_{13}=C_{24}=0$. These are shown in Figs. 2(a) and 2(b).
Although the entanglement of three and four qubits cannot be
discriminated, the additional entanglement of three and four
qubits can be described by $1-\sum C_{ij}^2 =0.16$ [18, 19]. When
$0.41\leq B\leq1.0$, $IC_1=IC_2=IC_3=IC_4=\sqrt 3/2$. The global
entanglement $Q$ and the two-qubit concurrence $C$ satisfy the
relation of $Q=\frac12\sum_{ij}C^2_{ij}$ [18, 19]. There is no
additional entanglement of three and four qubits. When $B>1.0$,
there is no entanglement at all. All the values of $C$, $Q$ and
$IC_i$ equal to zero. It is very interesting to note that in the
low temperature limit the entanglement between two alternate
qubits of the four-qubit $XXM$ Heisenberg model undergoes two
sudden changes when the magnetic field $B$ is increased. This
novel property may be used as quantum entanglement switch in
quantum computing and quantum communications.

For a more general model of four-qubit Heisenberg $XX$ chain, the
anisotropic contribution needs to be considered. If the anisotropy
is included, the Hamiltonian of the system can be written as
\begin{equation}
\label{eq14}
H_{XXZM}=H_{XXM}+\frac{J\Delta}2\sum_{n=1}^4\sigma_n^z\sigma_{n+1}^z
\end{equation}
where $\Delta$ is the anisotropy parameter, $H_{XXM}$ is given by
Eq. (4). The system reduces to $XX$ model when $\Delta=0$ and the
isotropic $XXX$ model when $\Delta=1.0$.

The concurrence $C$ of alternate qubits is plotted in Fig. 3 as
functions of the temperature $T$, the magnetic field $B$, and the
anisotropic parameter $\Delta$ when the strength of interaction
$J$ is $1.0$. The concurrence $C$ is plotted as functions of $T$
and $\Delta$ in Fig. 3(a) when $B=0.5$. From Fig. 3(a), it is seen
that the concurrence $C$ is increased first, then reached a
maximum value, and finally decreased when $\Delta$ is increased
from $-0.4$ to $1.0$. For $\Delta> 0$, the concurrence $C$ is
monotonically decreased when $T$ is increased. The concurrence $C$
is plotted as functions of $B$ and $\Delta$ in Fig. 3(b) when
$T=0.2$. From Fig. 3(b), it is seen that the value of $C$ is kept
zero for $\Delta\geq 0$ when $B=0$. The peak in $C$ appears at
$\Delta=-0.5$. It is found that the anisotropy in the Heisenberg
model can induce the entanglement between alternate qubits even
for $B=0$. If $B>0$, the height of the peak in $C$ is increased
and the position of the peak is shifted to larger values of both
$B$ and $\Delta$. From Fig. 3, it is seen that the curve of $C$ is
asymmetric about $\Delta$.

In conclusion, in this report the entanglement between two
alternate qubits of a four-qubit Heisenberg $XX$ model is
investigated. There is no thermal entanglement between alternate
qubits of a four-qubit Heisenberg $XX$ model when $B$ is very
small. However, when the magnetic field $|B|$ is increased, it can
induce entanglement in the $XX$ model both for the
antiferromagnetic and ferromagnetic cases. The square wave like
shape appeared in the concurrence $C$ may be used as quantum
entanglement switch. It is very interesting to find that the
temperature affects the entanglement much stronger for weakly
interacting alternate qubits than that for strongly interacting
nearest qubits. The anisotropy in the Heisenberg model can also
induce entanglement between alternate qubits.

\bigskip

\textbf{Acknowledgement}

It is a pleasure to thank Yinsheng Ling, Jianxing Fang, and Xiang
Hao for their many helpful discussions and calculations.

\newpage

\textbf{REFERENCES}

[1] C. H. Bennett, G. Brassard, C. Crepeau, R. Jozsa, A. Peres,
and K. Wooters,

\ \ \ \ Phys. Rev. Lett. {\bf 70}, 1895 (1993).

[2] A. K. Ekert, Phys. Rev. Lett. {\bf 67}, 661 (1991).

[3] M. Murao, D. Jonathan, M. B. Plenio, and V. Vedral, Phys. Rev.
{\bf A59},

\ \ \ \ 156 (1999).

[4] B. E. Kane, Nature {\bf 393}, 133 (1998).

[5] C. H. Bennett and D. P. Divincenzo, Nature {\bf 404}, 247
(2000).

[6] M. C. Arnesen, S. Bose, and V. Vedral, Phys. Rev. Lett. {\bf
87}, 017901 (2001).

[7] X. G. Wang, Phys. Rev. {\bf A64}, 012313 (2001).

[8] G. L. Kamta and A. F. Starace, Phys. Rev. Lett. {\bf 88},
107901 (2002).

[9] Y. Sun, Y. G. Chen, and H. Chen, Phys. Rev. {\bf A68}, 044301
(2003).

[10] L. F. Santos, G. Rigolin and C. O. Escobar, Phys. Rev. {\bf
A69}, 042304 (2004).

[11] X. G. Wang, H. Fu, and A. I. Solomon, J. Phys. {\bf A34},
11307 (2001).

[12] X. G. Wang, Phys. Rev. {\bf A66}, 034302 (2002).

[13] X. Q. Xi, W. X. Chen, S. R. Hao, and R. H. Yue, Phys. Lett.
{\bf A300},

\ \ \ \ \ \ 567 (2002).

[14] S. Hill and W. K. Wootters, Phys. Rev. Lett. {\bf 78}, 5022
(1997).

[15] W. K. Wootters, Phys. Rev. Lett. {\bf 80}, 2245 (1998).

[16] M. A. Nielsen, Ph. D. thesis, University of New Mexico, 1998,
quant-ph/0011036.

[17] M. C. Arnesen, S. Bose, and V. Vedral, Phys. Rev. Lett. {\bf
87}, 017901 (2001).

[18] J. Endrejat and H. B\"{u}ttner, quant-ph/0406006.

[19] U. Glaser, H. B\"{u}ttner, and H. Fehske, Phys. Rev. {\bf
A68}, 032318 (2003),

\ \ \ \ \ \ quant-ph/0305108.

[20] G. Brennen, QIC {\bf3}, 619(2003), quant-ph/0305094.

\newpage

\textbf{FIGURE CAPTIONS}

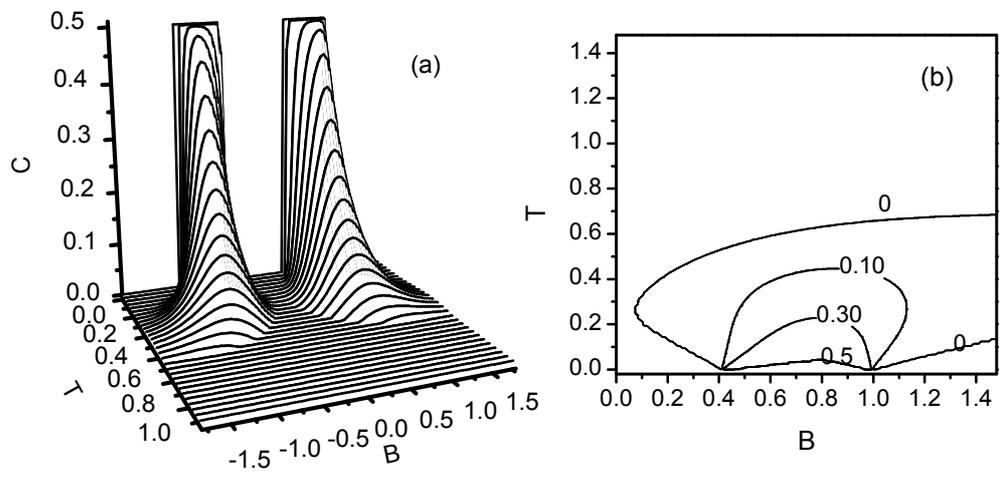
{\bf Fig. 1.}

The concurrence $C$ is plotted as functions of magnetic field $B$
and temperature $T$ when $J=1.0$.

(a). The curve of $C$ as functions of $B$ and $T$.

(b). The contour map of $C$ as functions of $B$ and $T$.

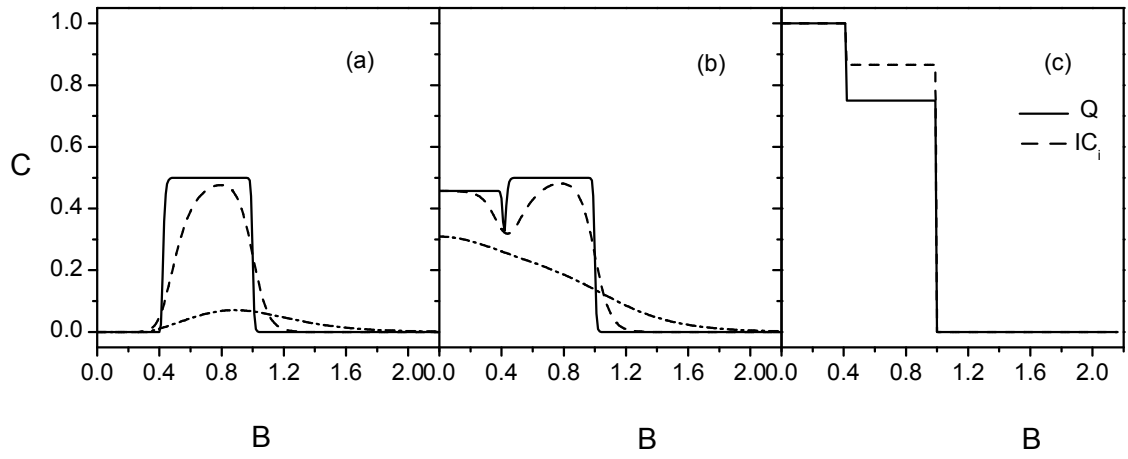
{\bf Fig. 2.}

The concurrence $C$, the global entanglement $Q$, and the
$i$-concurrence $IC_i$ are plotted as a function of $B$ when
$J=1.0$. For the concurrence $C$ of (a) and (b), the curve is
plotted when $T=0.01, 0.1$, and $0.5$ (from top to bottom).

(a). The curve $C$ of the alternate qubits.

(b). The curve $C$ of the nearest-neighbor qubits.

(c). The curves of $Q$ and $IC_i$ are plotted when $T=0.01$.

     ------ : The curve of $Q$; - - - : The curve of $IC_i$.

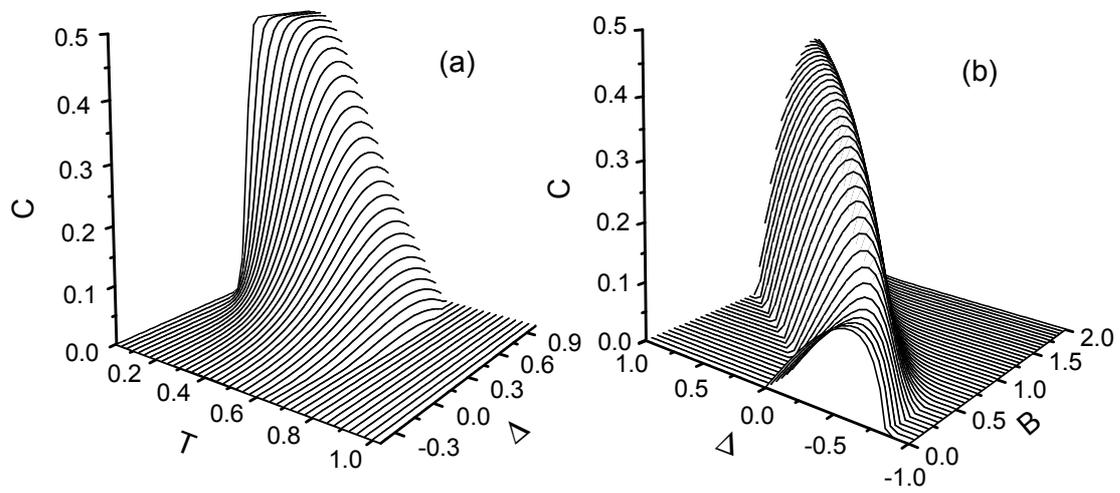
{\bf Fig. 3.} The concurrence $C$ between alternate qubits is
plotted as functions of $T$, $B$, and $\Delta$ when $J=1.0$.

(a). The concurrence $C$ as functions of $T$ and $\Delta$ when
$B=0.5$.

(b). The concurrence $C$ as functions of $B$ and $\Delta$ when
$T=0.2$.

\end{document}